\documentclass[twocolumn,showpacs,aps,prb]{revtex4-1}

\usepackage{amssymb,amsmath}

\usepackage{graphicx}
\usepackage{color}
\begin{document}

\title{Compressibility as a probe of
quantum phase transitions\\
 in   topological superconductors}

\author{David Nozadze}
\author{Nandini Trivedi}
\affiliation{Department of Physics, The Ohio State University, 191 W. Woodruff Avenue, Columbus, OH 43210, USA}

\begin{abstract}
The non-Abelian statistics of Majorana fermions, their role in topological quantum computation, and the possibility of realizing them in condensed matter systems, has attracted considerable attention. While there have been recent reports of zero energy modes in single particle tunneling density of states, their identity as Majorana modes has  so far not been unequivocally established. We make predictions for the local compressibility $\kappa_{\rm{loc}}$,  tuned by changing the chemical potential $\mu$ in a semiconducting nanowire with strong spin-orbit coupling and in a Zeeman field in proximity to a superconductor, that has been proposed as a candidate system for observing Majorana modes. 
We show that in the center of the wire, the topological phase transition is signaled by a divergence of  $\kappa_{\rm{loc}}$  as a function of $\mu$ which is an important diagnostic of the topological phase transition.
We also find that a single strong impurity potential can lead to a local {\it negative} compressibility at the topological phase transition. The origin of such anomalous behavior can be traced to the formation of 
Andreev bound states close to topological phase transitions. 
Measurable by a scanning electron transistor, the compressibility includes contributions from both single particle states and collective modes and is therefore a complimentary probe 
from scanning tunneling spectroscopy.

\end{abstract}

\date{\today}

\maketitle

%%%%%%%%%%%%%%%%%%%%%%%%%%%%%%%%%%%%%%%%%%%%%%%%%%%%%%%%%%%%%%%%%%%%%%%%%%%%%%%%%
% Main text starts here
%%%%%%%%%%%%%%%%%%%%%%%%%%%%%%%%%%%%%%%%%%%%%%%%%%%%%%%%%%%%%%%%%%%%%%%%%%%%%%%%%

%%%%%%%%%%%%%%%%%%%%%%%%%%%%%%%%%%%%%%%%%%%%%%%%%%%%%%%%%%%%%%%%%%%%%%%%%%%%%%%%%
The search for  Majorana fermions  in condensed matter systems and their 
possible application for topological quantum computation \cite{2008_Nayak_RMP,  2009_Wilczek_NP,  2003_Kitaev_AP,   1991_Moore_NPB, 2012_Alicea_RPP}
has led to several
promising proposals for practical  realizations of Majorana fermions  both in one (1D) and two-dimensional (2D) systems.
Majorana fermions can emerge in  systems, such as topological insulator-superconductor interfaces \cite{2008_Fu__PRL,  2010_Linder_PRL}, quantum Hall
states with filling factor~\cite{1991_Moore_NPB} 5/2, $p$-wave superconductors \cite{2000_Read_PRB}, semiconductor heterostructures \cite{2010_Sau_PRL, 2010_Alicea_PRB},
half-metallic ferromagnets \cite{2011_Duckheim_PRB, 2011_Chung_PRB} and ferromagnetic metallic chains \cite{2012_Potter_PRB}. As shown by Kitaev \cite{2001_Kitaev_PU}, Majorana fermions can emerge at the ends of 1D spinless $p$-wave superconducting chain when 
the chemical potential is in the topological regime. 

A realization of the Kitaev chain based on
a quantum nanowire made of a semiconductor-superconductor hybrid structure has been proposed \cite{2010_Sau_PRL, 2010_Alicea_PRB}. 
In the presence of Rashba spin-orbit coupling, the parabolic bands for the two spin projections get separated. In addition, a Zeeman field $h$ opens up a gap leading to an effectively  spinless 1D system when
the chemical potential $\mu$ lies in the Zeeman gap. The proximity induced superconductivity with a gap  $\Delta$ can result in the  topological phase  \cite{2010_Sau_PRL, 2010_Alicea_PRB, 2010_Lutchyn_PRL, 2012_Mourik_S}. In this regime, the wire can be realized as a Kitaev chain
and should have two Majorana localized zero energy modes at the ends. The nanowire can undergo a quantum phase transition from  a topologically trivial superconducting phase to the topological one (or vise verse) by changing  the chemical potential or the  magnetic field. 

There have been recent reports of observations of Majorana fermions in tunneling and
the fractional Josephson effect \cite{2001_Sengupta_PRB, 2004_Kwon_LTP, 2012_Mourik_S, 2012_Rokhinson_NP}. 
Ref.~\onlinecite{2014_Yazdani_Sc} has reported significant progress in creating the Majorana states where  spatial location of Majoranas are detected using a scanning
 tunneling microscope.  All these experimental observations of the existence of Majorana fermions {\em assume} that the system is in the topological phase and attribute the zero energy density of states to the proposed Majorana modes. However, since there could be several other sources of zero bias anomaly \cite{2000_Sasaki_Nature, 2002_Zareyan_PRB, 2013_Sau_PRB}, the existence of the Majorana modes has so far not been unequivocally established. 
 
\begin{figure*}[!htb]
\includegraphics[width=17cm]{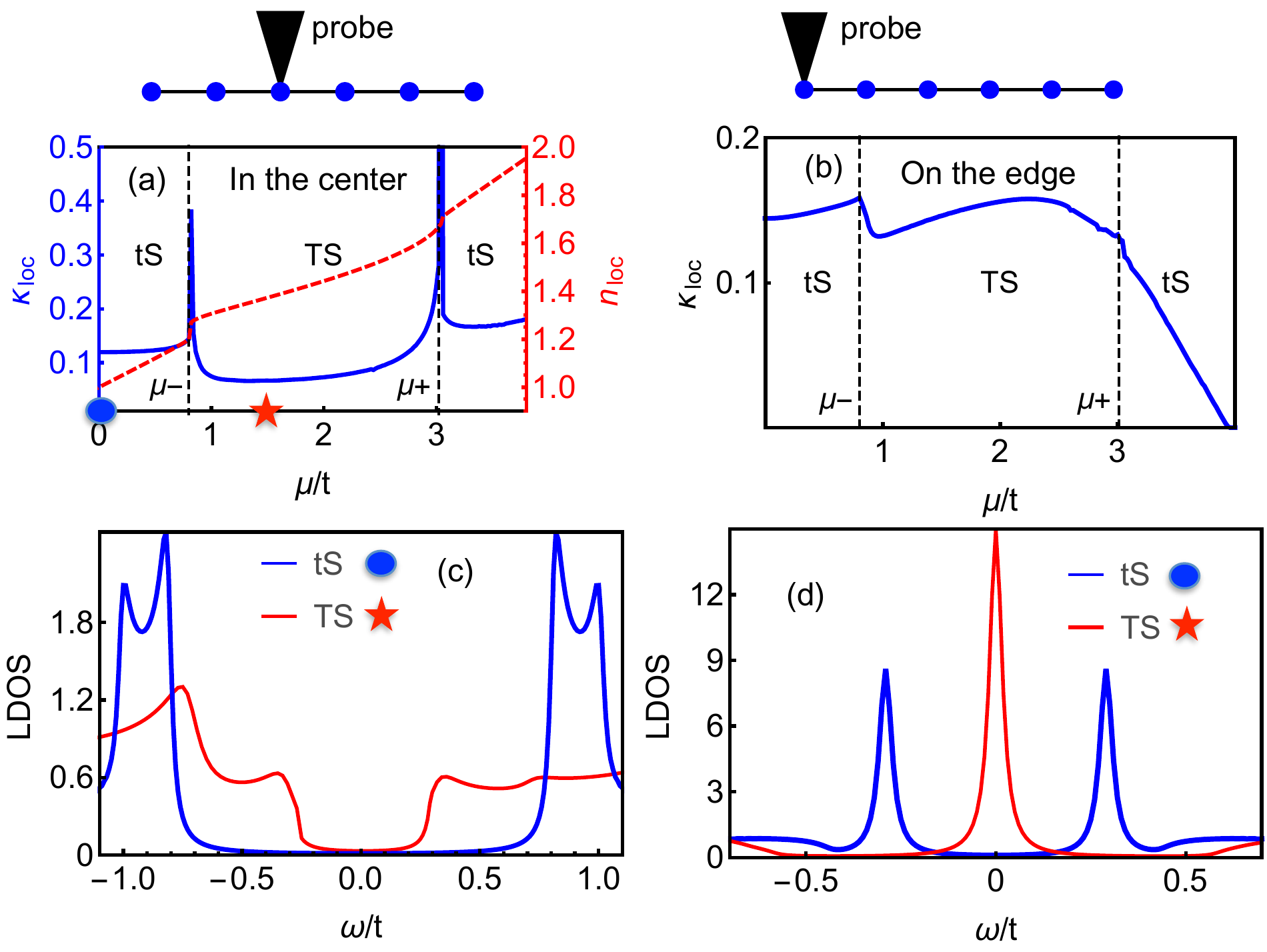}
\caption{(Color online). (a) The local particle density  $n_{\rm{loc}}$ (dashed line) and the local compressibility $\kappa_{\rm{loc}}$  (solid line) versus  chemical potential $\mu/t$ in the center of wire. 
The compressibility  has  sharp peaks  at the transition between trivial (tS) and topological (TS) superconducting phases.
The singularity in $\kappa_{\rm{loc}}$ is weakened at the edge of the wire (b).  (c,d) The local density of states $N_{\rm{loc}}(\omega)$, measured relative to the chemical potential $\mu$,  in the trivial ($\mu=0$, blue circle) and topological ($\mu=1.5t$, red star) phases. Results presented for chain length $N = 256$ at $T=0$ with slightly broadened $\delta$-functions in  $N_{\rm{loc}}(\omega)$.}
\label{fig:fig1}
\end{figure*}

We propose here a definitive method to determine whether or not the nanowire is in the topological or trivial state through measurements of the gate-tuned local compressibility. 
We find very different behavior of the local compressibility at the edge of the wire and in the center. While the edge harbors Majorana modes which show a zero bias anomaly, the 
density of states in the center is fully gapped. The compressibility on the other hand 
shows a sharp singularity in the center of the wire at the topological to trivial phase transitions (see Fig. \ref{fig:fig1}). This is one of our central results.
Another key result is that a single impurity can dramatically change the local response: a strong impurity leads to the formation of Andreev bound states  and surprisingly this results in a local negative compressibility with a dip at the topological phase transitions as the chemical potential is tuned.  An extra peak associated with the bound state appears in the local 
compressibility above the transition. 

\begin{figure*}[!htb]
\includegraphics[width=17cm]{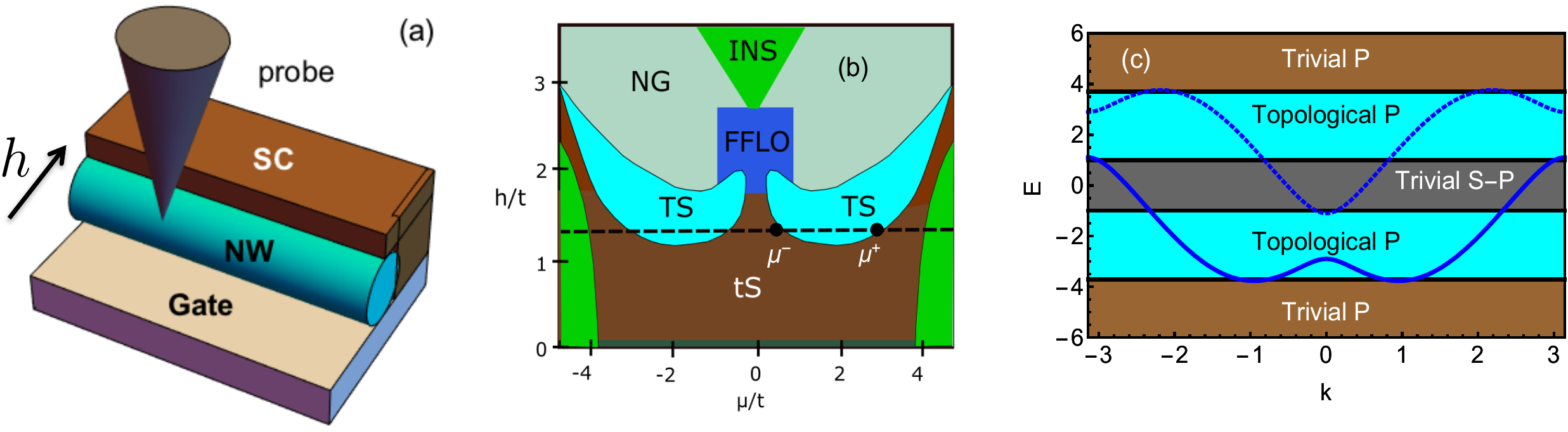}
\caption{(Color online). (a) Proposal for measuring local compressibility using an SET (single electron transistor) in the suggested set-up\cite{2010_Sau_PRL, 2010_Alicea_PRB} 
of a nanowire (NW) with spin-orbit coupling 
in proximity to a superconductor (SC) with an applied magnetic field.  (b) Phase diagram of 1D spin-orbit coupled superconductor as function of Zeeman field $h/t$
and the chemical potential $\mu/t$~\onlinecite{2014_QU_PRA}.  Five different phases can be identified: trivial superconducting (tS), topological superconducting (TS), FFLO, normal gas (NG) and insulator phase (INS).   (c) The band structure for a wire with spin-orbit coupling in a magnetic field. As attraction is turned on, different pairing symmetries emerge depending on the location of $\mu$ 
(the Bogoliubov bands have not been shown). Within the first band, 
the system is described by the Kitaev model that captures the transition from the trivial $p$ to topological $p$-wave SC. Once the second band is crossed, 
both interband $s$-wave and intraband $p-$ wave channels become operative. }
\label{fig:fig2}
\end{figure*}

\begin{figure*}[!htb]
\includegraphics[width=17cm]{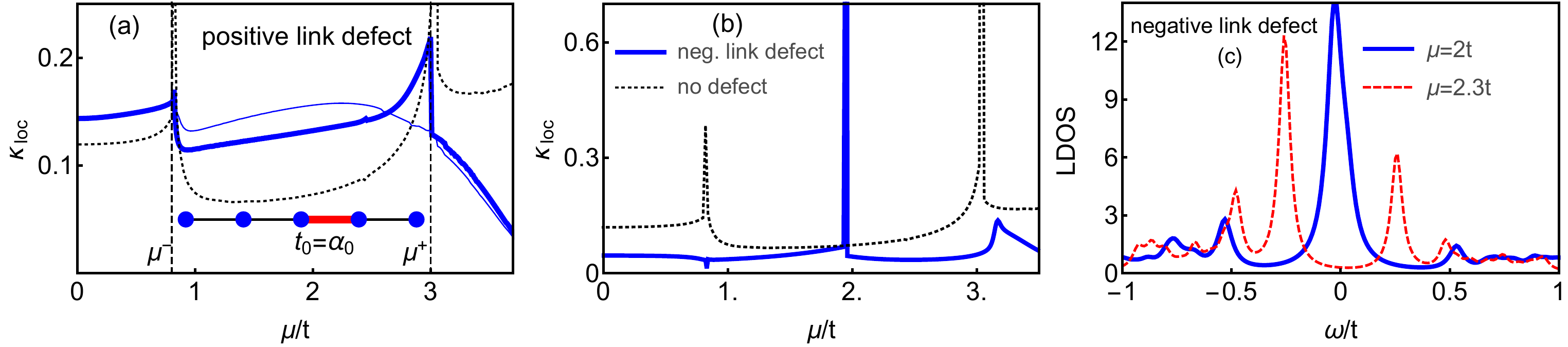}
\caption{(Color online). Panels (a,b) The local compressibility $\kappa_{\rm{loc}}$ as a function of the chemical potential $\mu/t$ for link defects measured on either side of the link. 
Panel (a) Positive link defect (spin-orbit coupling  
$\alpha_0\neq \alpha$ and hopping parameter $t_0\neq t$ different from the reference values shown for three cases (i) No defect (black dotted line); (ii) $\alpha_0=t_0=0.3t$ (blue thick solid line); 
(iii)  cut wire with $\alpha_0=t_0=0$ (blue thin line). The singularity in $\kappa_{\rm{loc}}$ at the topological phase transitions at $\mu=\mu^{\pm}$ is weakened in the presence of the link defect. 
Panel (b) Negative link defect with $t_0=-t$ (blue solid line).
The sharp peak in $\kappa_{\rm{loc}}$ within the topological phase arises because of the formation of a zero-energy bound state on either side of the link defect. 
(c) Local density of states (LDOS) for a negative link defect showing a bound state at zero energy at $\mu=2t$ (blue) and away from zero $\mu=2.3t$ (red).}
\label{fig:fig3}
\end{figure*}

\begin{figure*}[!htb]
\includegraphics[width=17cm]{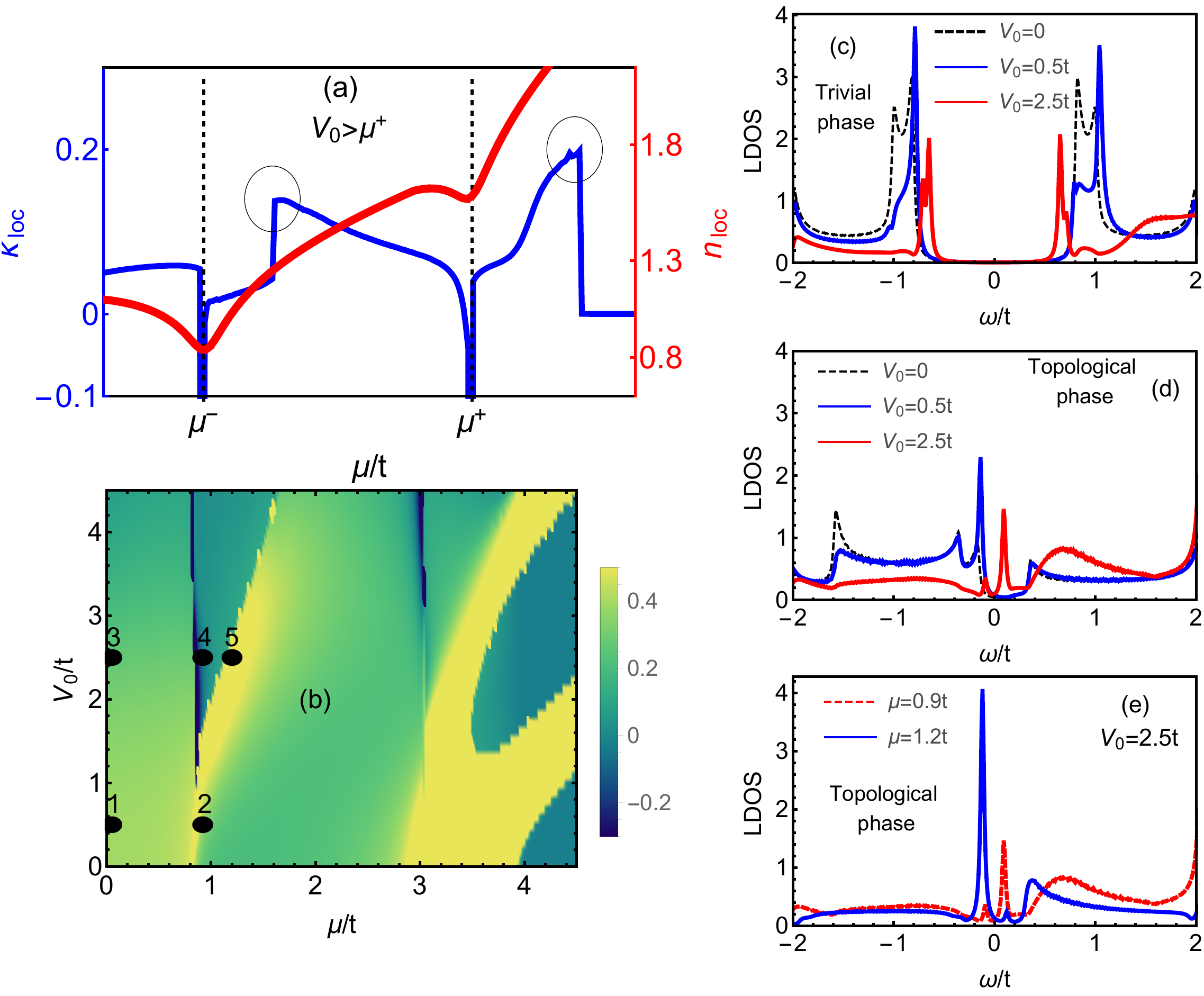}
\caption{(Color online). 
(a) The local particle density $n_{\rm{loc}}$ and $\kappa_{\rm{loc}}$ for a local potential  defect $V_0>\mu^+$. Close to the topological phase transitions $\mu=\mu^\pm$, $\kappa_{\rm{loc}}$
becomes negative. In addition, extra peaks (shown by circles) appear in the topological and trivial phases.  
(b) Density plot of $\kappa_{\rm{loc}}$ in the $\mu/t$ -- $V_0/t$ plane.  
(c,d,e) Local density of states for various values  of $V_0$ and $\mu$.
(c) shows how a in-gap state appears in the presence of an on-site impurity in the trivial phase ($\mu=0<\mu^-$) with  $s$ and $p$ wave parings.  
(d) shows how a non-trivial bound state forms in the topological phase close to the topological  phase transition at $\mu=0.9t>\mu^-$ as impurity strength $V_0$ increases. 
The bound state starts to be formed when $V_0>\mu^-$. (e)  The bound state in the topological phase for $\mu=1.2t$. 
}
\label{fig:fig4}
\end{figure*}

%%%%%%%%%%%%%%%%%%%%%%%%%%%%%%%%%%%%%%%%%%%%%%%%%%%%%%%%%%%%%%%%%%%%%%%%%%%%
\noindent {\it Model and methods:}
%%%%%%%%%%%%%%%%%%%%%%%%%%%%%%%%%%%%%%%%%%%%%%%%%%%%%%%%%%%%%%%%%%%%%%%%%%%%%%
We consider a 1D tight-binding Hamiltonian 

\begin{align}\label{ham0}
H&=-\sum_{i,\sigma}(\mu-V_i) c^\dagger_{i\sigma}c_{i\sigma}-\sum_{i\sigma}t_i(c^\dagger_{i\sigma}c_{i+1\sigma}+{\rm{h.c.}})\nonumber\\
&+H_{\rm{SO}}+H_{\rm{Z}}+H_{\rm{Int}}\,,
\end{align}
where $c^\dagger_i$ $(c_i)$ is the creation (destruction) operator for an electron
on a site $i$, $t_i$ the nearest-neighbor hopping and  $\mu$ is the chemical potential.
The spin-orbit coupling  and the Zeeman field terms are given by
$H_{\rm{SO}}=\frac{1}{2}\sum_{i\sigma}\alpha_i\left[c^\dagger_{i+1\sigma}
(i\sigma_y)_{\sigma\sigma'}c_{i\sigma'}+{\rm{h.c.}}\right]$ and 
$H_{\rm{Z}}=-h\sum_{i\sigma}c^\dagger_{i+1\sigma}(\sigma_z)_{\sigma\sigma'}c_{i\sigma'}$, respectively.
Parameters $\alpha_i$ refer to the Rashba spin-orbit coupling and  $h$ to the Zeeman field.  $V_i$ is the on-site impurity potential. The interaction term 
$H_{\rm{Int}}=-U\sum_ic^\dagger_{i\uparrow}c_{i\uparrow}c^\dagger_{i\downarrow}c_{i\downarrow}$, where $U$ is the pairing interaction.
In the clean limit $\alpha_i=\alpha$, $t_i=t$ and $V_i=0$.

\medskip
We solve the model in Eq.~(\ref{ham0}) within the Bogoliubov-de Gennes (BdG) self-consistent  approach and calculate the local particle density $n_{\rm{loc}}$ and
the local density of states $N_{\rm{loc}}(\omega)$; (see Supplement for more details).
Even though this is a one-dimensional problem we are justified in ignoring the quantum fluctuations, primarily because the system is proximity coupled to a bulk superconductor which damps out the fluctuations.  

As shown in Ref.~\onlinecite{2014_QU_PRA}, the model in Eq.~(\ref{ham0}) has several different phases: trivial superconducting phase (tS), topological superconducting phase (TS) with Majorana fermions at ends of chain, Fulde-Ferrell-Larkin-Ovchinnikov phase (FFLO) with spatially oscillating order parameter $\Delta$ and non-zero magnetization,  insulator phase (INS) with finite energy gap and normal gas (NG) phase without pairing and energy gap (see Fig.~\ref{fig:fig2}  (b)).  
We are interested in a particular slice of the phase diagram in order to investigate the behavior of the compressibility across the  topological to trivial phase transitions.  We consider the Zeeman field $h=1.35t$ and spin-orbit coupling $\alpha=t$ that shows two transitions from the topological to the trivial phases as a function of the chemical potential $\mu$ at 
$\mu^{-}$ and $\mu^{+}$. 
We find the quasiparticle excitation energy $E^2_{\pm}(k)=
\epsilon_k^2+\alpha^2\sin^2(k)+h^2+\Delta^2\pm 2\sqrt{h^2(\epsilon^2_k+\Delta^2)+\alpha^2\sin^2(k)\epsilon^2_k}$, where $\epsilon_k=-2t\cos(k)-\mu$. When $h>0$, the gap ($E_{-}=0$) closes at
$|\mu^{\pm}|=(2t\pm\sqrt{h^2-\Delta^2})$,  which corresponds to the phase transitions. The system is in the  topological phase when $\mu^-<|\mu|<\mu^+$ and  in the trivial phase for $\mu<-\mu^+$ or $\mu>\mu^+$  (see Fig.~\ref{fig:fig2} (c)). For the parameters chosen, 
$\mu^-\approx0.84t$ and $\mu^+\approx 3t$. It is important to note that the system is effectively a ``spinless" 
$p$-wave superconductor so long as the chemical potential crosses only a single band and the transitions are from the topological phase to a trivial $p$-wave phase. Once both bands are crossed, the trivial phase has contributions from both $s$ (inter-band) and $p$-wave (intra-band) paring channels\cite{2012_Alicea_RPP}.

\medskip

\noindent{\it{Compressibility at the topological phase transition:}}
From the self consistent BdG solutions of Eq.~(\ref{ham0}), we obtain the local particle number $n_{\rm{loc}}$ and its dependence on the global chemical potential $\mu$
yields the compressibility $\kappa_{\rm{loc}}=\partial n_{\rm{loc}}/\partial \mu$. We find that in the center of the wire the local compressibility shows a logarithmic divergence 
arising from the gap closing linearly between the topological to the trivial phase transitions (Fig.~{\ref{fig:fig1}} (a)). In contrast, the singularities are weakened on the edge of the wire (Fig.~{\ref{fig:fig1}} (b)).
It is useful to contrast the compressibility which captures the single particle and the pair (collective modes) density of states (DOS), from the behavior of  the single particle density of states $N_{\rm{loc}}(\omega)$. 
As seen in Fig.~\ref{fig:fig1} (c,d), the DOS shows a gap in both the topological and the trivial phases except at the transition where the gap gets closed. However, the compressibility  is  non-zero inspite of a single-particle gap because of the contribution from the pairs. 

This is one of our central results. It highlights the fact that by measuring both the local tunneling DOS at the edge of the wire {\em and}
the local compressibility at the center of the wire as a function of $\mu$ it is possible to unequivocally determine when the wire is in the topological phase with Majorana modes localized at the edges.
As a control, $\mu$ can be varied to bring the wire into a trivial phase with a finite compressibility and a gapped single particle DOS. 

\medskip

We next discuss the effect of a weak link and a single potential disorder on the local compressibility.

\noindent{\it {Weak link}}: In the presence of a weak link, defined by a hopping $t_0\neq t$, the local particle number $n_{\rm{loc}}$ becomes inhomogeneous.
We calculate the local compressibility on either side of the link defect $\kappa_{\rm{loc}}=\partial n_{\rm{loc}}/\partial \mu$ by differentiating it with respect to the global chemical potential $\mu$. 

For a positive link defect, i.e. $t_0$ has the same sign as $t$, we find that the peaks in the local compressibility are weakened at the transition point (Fig.~\ref{fig:fig3} (a)).  In the limit $t_0=0$ the wire is cut, the singularity in $\kappa_{\rm {loc}}$ is completely suppressed, though the compressibility  remains finite.
We also consider the case of a link with negative hopping parameter i.e.
$t_0=-t$.  Such a negative link can be produced by a local $\pi$-junction. The behavior of $\kappa_{\rm {loc}}$ as a function of $\mu$ is remarkably different from the positive defect. In the topological phase 
a sharp peak appears in the local compressibility measured on the either side of the link (Fig.~\ref{fig:fig3} (b)) at a particular value of $\mu_0$; for the chosen parameters, $\mu_0=2t$.
This is due to the formation of a zero-energy bound state (Fig.~\ref{fig:fig3} (c)) at $\mu_0$. It is important to note that this zero-energy bound state is formed only in the topological phase.
At the same time, it does not correspond to a Majorana mode based on the structure and symmetries of the corresponding eigenfunctions (see Supplement for more details). For $\mu\neq \mu_0$ 
the bound state moves away from zero energy and no longer contributes to the singularity in the compressibility.

\noindent{\it  {Local potential}}:
In the presence of an on-site impurity $V_0$, there are several interesting features in the behavior of the local particle density and compressibility as shown in the density plot in Fig.~\ref{fig:fig4} (b).

\noindent {\it{(i) Negative local compressibility}}: 
A repulsive potential  $V_0\gtrsim \mu^{-} $ can have a non-trivial effect on the local density and compressibility.
The local particle density is found to {\it decrease} around the topological phase transition at $\mu=\mu^-$ 
even as $\mu$ increases. Correspondingly,  the  local compressibility  $\kappa_{\rm{loc}}$ becomes negative  and shows a dip at the transition.  
For impurity strength somewhat larger  than $V_0\gtrsim \mu^{+}$, in addition to the dip at $\mu=\mu^-$, a second dip appears 
at $\mu=\mu^{+}$ where also the  local compressibility  $\kappa_{\rm{loc}}$ becomes negative (Fig.~\ref{fig:fig4} (a)). The plot for the particle density 
shown here is schematic for clarity, the actual data with more details can be found in the supplementary information. 
The reason for the decrease of the local density and the corresponding negative local compressibility is tied to the formation of an impurity bound state (BS) above zero energy that  starts to form close to the topological phase transitions.

\noindent{\it{(ii) Bound states}}: 
As seen in Fig.~\ref{fig:fig4} (c), In the presence of an on-site impurity, the peaks at the gap edge are suppressed and the gap size is reduced. This can be understood from the fact that
the trivial phase has both $s$ and $p$-wave paring, and disorder affects the $p$-wave component more drastically than the $s$-wave component; however the spectrum remains gapped.
In the topological phase where the system is effectively a ``spinless''  unconventional ($p$-wave) superconductor, a bound state is formed due to the sign change of the order parameter in this unconventional superconductor. 

Figure \ref{fig:fig4} (d) shows
how the zero-energy bound state starts to form when $V_0\gtrsim \mu^-$. 
For a fixed $V_0$ as $\mu$ increases,  the bound state  becomes sharper and moves to zero energy. 
At this point the zero-energy bound state is detectable as an additional feature shown by a circle in $\kappa_{\rm{loc}}$ (Fig.~\ref{fig:fig4} (a)). With further increase of $\mu$ the BS moves below 
the chemical potential (Fig.~\ref{fig:fig4} (e)).
Similarly, a zero-energy  BS forms also in the trivial $p$-wave phase for the impurity strength  $V_0>\mu^+$. 

For a negative impurity potential, the BS  forms below the Fermi level and more states shift below the Fermi energy to enhance the local density for all $\mu$. 
In contrast to the scenario of the positive impurity potential, the BS does contribute to the local particle density for a negative impurity. As the result, the local particle density starts to increase as 
$\mu$ decreases, until a sharp BS is formed. This once again causes the local compressibility to become negative around the topological phase transition; (see Supplement for more details).

%%%%%%%%%%%%%%%%%%%%%%%%%%%%%%%%%%%%%%%%%%%%%%%%%%%%%%%%%%%%%%%%%%%%%%%%%%%%%%%%%

\medskip

\noindent{\it {Conclusions:}}
Our theoretical proposals based on the compressibility, in conjunction with scanning tunneling spectroscopy, are powerful diagnostics for detecting topological phase transitions in 1D spin-orbit coupled superconductors. Specifically in the presence of local defects,  the local compressibility can be measured
using single-electron transistor (SET) spectroscopies  \cite{2008_Martin_NatPhys}. Ref.~\onlinecite{2008_Martin_NatPhys} has in fact used the SET in a different context to measure the inverse compressibility locally on a graphene sample as a function of the back-gate voltage or carrier density.  We expect the same technique can be applied to the spin-orbit coupled nanowires - superconductor devices to detect the topological phase transition
guided by our predictions.

Some of the most promising directions to experimentally investigate are: (a) the sharp  peak in the compressibility at the topological phase transition tuned by the Zeeman field in the clean wire, and,
(b) the negative compressibility induced by the on-site impurity in the topological phase.
In general it will be useful to see the interplay between local scanning and local compressibility spectroscopies for giving insights into single particle and collective modes.

%%%%%%%%%%%%%%%%%%%%%%%%%%%%%%%%%%%%%%%%%%%%%%%%%%%%%%%%%%%%%%%%%%%%%%%%%%%%%%%%%

%%%%%%%%%%%%%%%%%%%%%%%%%%%%%%%%%%%%%%%%%%%%%%%%%%%%%%%%%%%%%%%%%%%%%%%%%%%%%%%%%
%\appendix*
%%%%%%%%%%%%%%%%%%%%%%%%%%%%%%%%%%%%%%%%%%%%%%%%%%%%%%%%%%%%%%%%%%%%%%%%%%%%%%%%%

\section*{Acknowledgements}
DN and NT were supported by the NSF under Grant No. NSF-DMR1309461.

\bibliographystyle{apsrev4-1}

%

%\bibliography{/Users/davide/Dropbox/b}
%\bibliography{/Users/nandinitrivedi/Dropbox/b}
%\bibliography{/Users/david/Dropbox/b}
%\bibliography{b}

\end{document}

\title{Supplementary Information for: Compressibility as a probe of
quantum phase transitions\\
 in   topological superconductors} 
\author{David Nozadze}
\author{Nandini Trivedi}
\affiliation{Department of Physics, The Ohio State University, 191 W. Woodruff Avenue, Columbus, OH 43210, USA}
\begin{abstract}
\end{abstract}
\pacs{}
\maketitle

In this supplement, we present: (I) details of the Bogoliubov-de Gennes (BdG) mean field theory for the nanowire in proximity to a superconductor; 
(II) information about the eigenstates associated with zero-energy bound states that are localized around a negative link ($\pi$ phase shifted) in the topological phase but are {\it not} Majorana modes;
(III)  additional information about  the local particle density, local compressibility and the local density of states (LDOS) in the presence of an on-site impurity. 

\begin{figure*}
\includegraphics[width=17cm]{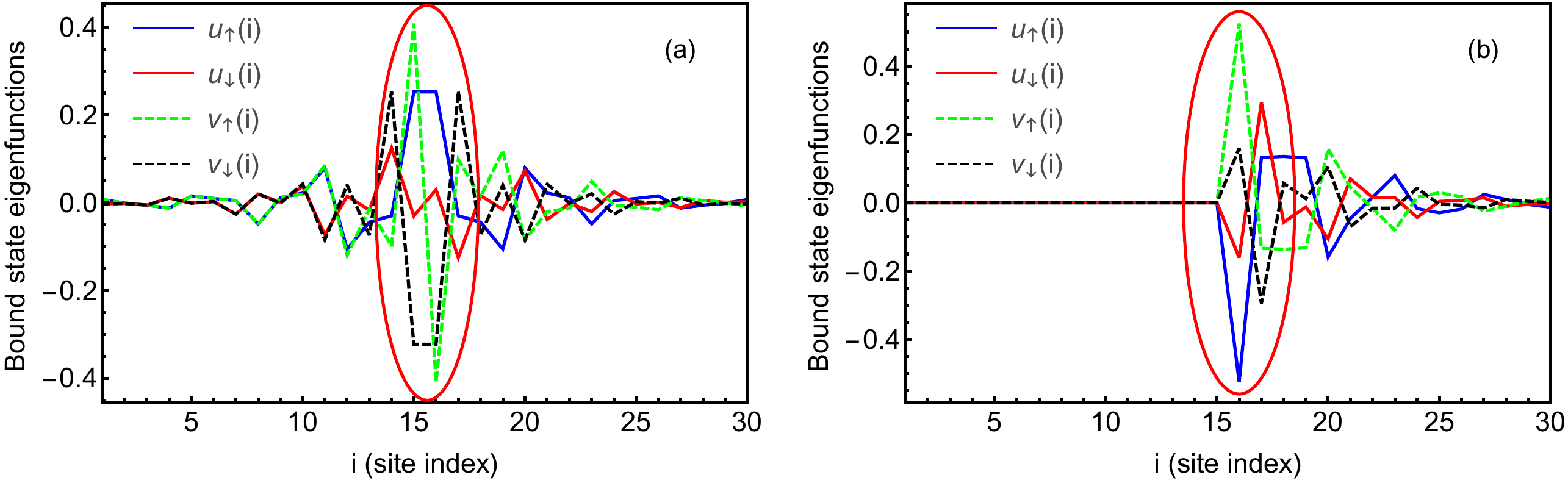}
\caption{(Color online). (a) The bound state eigenfunctions at zero-energy in the presence of a negative link impurity $t_0=-t$ between sites 15 and 16.
(b) The bound state eigenfunctions at zero-energy for a broken link $t_0=\alpha=0$ between sites 15 an 16.
From the symmetry conditions $u_\sigma(i)=-v_\sigma(i)$ required for a Majorana fermion, it is evident that while the bound state in (b) is a Majorana bound state that is topologically protected to be at zero energy, the bound state in (a) is an Andreev bound state that happens to be at zero energy.
The system size $N=256$, and parameters $\alpha=t$, $h=1.35t$ at $T=0$. }
\label{fig:fig3}
\end{figure*}

\begin{figure*}
\includegraphics[width=17cm]{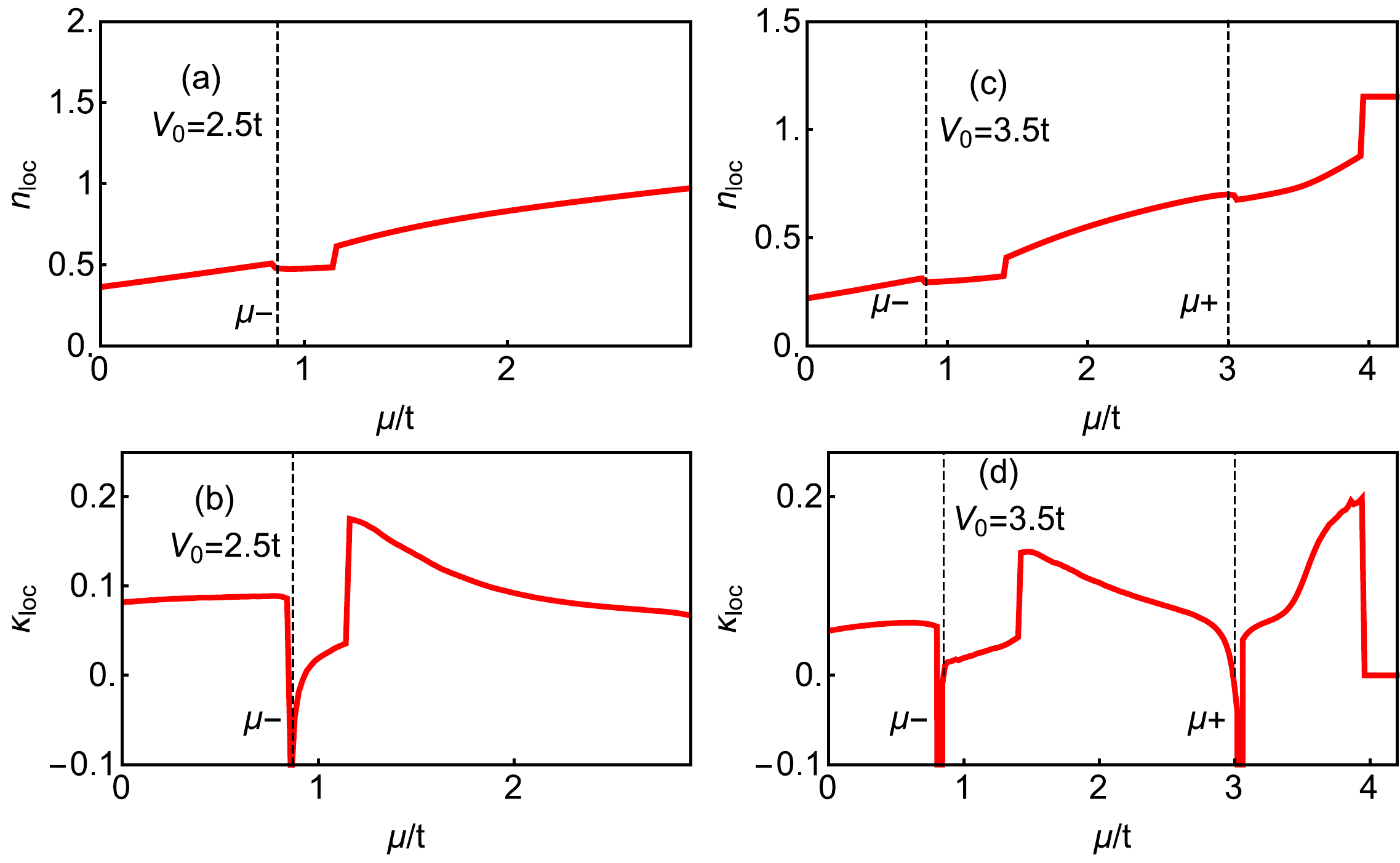}
\caption{(Color online). The local particle density  $n_{\rm{loc}}$ and the  compressibility $\kappa_{\rm{loc}}$  versus  chemical potential $\mu/t$ for two values of the on-site impurity potential $V_0$. }
\label{fig:fig1}
\end{figure*}

\medskip

\noindent {\bf I. Bogoliubov-de Gennes (BdG) approach:}

Starting with the full Hamiltonian in Eq. (1) in the main text, we perform a mean-field decomposition of the interaction term in the $s$-wave local pairing amplitude channel, 
$\Delta_{i}=-U\langle c^\dagger_{i\downarrow}c^\dagger_{i\uparrow}\rangle$. In addition, we can add an on-site impurity potential or a weak link. 
The mean field quadratic Hamiltonian is given by:
\begin{align}\label{ham1}
H&=-\sum_{i,\sigma}(\mu-V_i)c^\dagger_{i\sigma}c_{i\sigma}-\sum_{i\sigma}t_i(c^\dagger_{i\sigma}c_{i+1\sigma}+{\rm{h.c.}})\nonumber\\
&+H_{\rm{SO}}+H_{\rm{Z}}+\sum _i\Delta_i(c^\dagger_{i\uparrow}c^\dagger_{i\downarrow}+h.c.)\,.
\end{align}

We next diagonalize Eq.~(\ref{ham1}) using the Bogoliubov transformation $c_{i\sigma}=\sum_n \left(\gamma_n u_{n\sigma}(i)-\sigma \gamma^{\dagger}_nv_{n\sigma}(i)\right)$
which leads to BdG equations:
\begin{align}
\label{bdg}
 \left(\! \begin{array}{cccc}
\hat{H_\uparrow} & \hat{\alpha} & 0 & \hat{\Delta}\\
-\hat{\alpha} & \hat{H_\downarrow} &-\hat{\Delta}&0\\
0& -\hat{\Delta}^*&- \hat{H_\uparrow} &-\hat{\alpha}\\
\hat{\Delta}^*& 0&\hat{\alpha} &-\hat{H_\downarrow} \end{array} \!\right)\!\!
\left(\!\begin{array}{c}
u_{n\uparrow}(j)   \\
u_{n\downarrow}(j)  \\
-v_{n\uparrow}(j)\\
v_{n\downarrow}(j)  \end{array}\!\right)\!\!=\!\!E_n\!\!
\left(\!\begin{array}{c}
u_{n\uparrow}(j)   \\
u_{n\downarrow}(j)  \\
-v_{n\uparrow}(j)\\
v_{n\downarrow}(j)  \end{array}\!\right)
\end{align}
where the excitation eigenvalues $E_n\geq0$.  Here 
$\hat{H}_\sigma u_{n\sigma}(i)=(-\mu\pm h+V_i)u_{n\sigma}(i)-t_i(u_{n\sigma}(i+1)+u_{n\sigma}(i-1))$, $\hat{\alpha} u_{n\uparrow}(i)= \alpha(u_{n\downarrow}(i-1)-u_{n\downarrow}(i+1)) $ and $\hat{\Delta} u_{n\sigma}(i)=\Delta_{i}u_{n\sigma}(i)$. 
In terms of the eigenfunctions of the BdG hamiltonian, the local pairing amplitudes and local densities are given by:
\begin{align}
\Delta_i=-{U}\sum_{E_n\ge 0}  u_{n\uparrow}(i) v_{n\downarrow}(i)\tanh(E_n/2T).
\end{align} 
and
\begin{align}
\label{lnp1}
n_{i}=\frac{1}{2}\sum_{n,\sigma} \left[|u_{n\sigma}(i)|^2f(E_n)+|v_{n\sigma}(i)|^2(1-f(E_n))\right]\,,
\end{align}
where $f(E_n)$ is the Fermi function. 
The solutions of the BdG equations are iterated until self consistency is achieved at each site. In these studies we have set $T=0$. 
In the problem at hand, since the pairing in the nanowire is imposed by proximity with a bulk superconductor, it may be not necessary 
to obtain the pairing amplitudes self consistently, however in the presence of disorder
(site or link), the self consistency becomes more relevant and is included in our investigation.

Once self consistency is achieved  the single-particle density of states is obtained from
\begin{align}
\label{DOS1}
N_{i}(\omega)=\sum_{n,\sigma}\left[ |u_{n\sigma}(i)|^2\delta(\omega - E_n)+|v_{n\sigma}((i)|^2\delta(\omega + E_n)  \right]\,.
\end{align}
In the text the local density from Eq.~(\ref{lnp1}) is denoted by $n_{\rm loc}$ and the local single-particle density of states by $N_{\rm loc}(\omega)$.

\medskip

\myheading{II. Zero-energy bound state due to a weak link:}
As shown in the main text a negative link impurity $t_0=-t$ induces zero-energy bound states in the topological phase. Are these Majorana modes or simply Andreev bound states?
In order to get some insight about these modes we look at the corresponding eigenfunctions $u_\sigma(i)$ and $v_\sigma(i)$.   As seen in the Fig.~\ref{fig:fig3} (a), close to link impurity, the eigenstates do not satisfy the symmetry  conditions $u_\sigma (i)=v^\star_\sigma (i)$ or  $u_\sigma (i)=-v^\star_\sigma (i)$ that is required for a Majorana mode (Fig.~\ref{fig:fig3} (b)). Thus, even though these two  zero-modes  are in the topological phase these are not Majorana modes.

\medskip

\myheading{III. Local particle density and compressibility:} 
We present here the actual data in Fig.~\ref{fig:fig1} of the supplement for the local particle density and the local compressibility in the presence of an on-site impurity potential  $\mu^+> V_0 > \mu^-$, corresponding to Fig. 4 (a) of the main text. 
Close the topological phase transition at $\mu=\mu^-=0.84t$, the local particle density starts to decrease even as the
chemical potential increases and the local compressibility $\kappa_{\rm loc}$ becomes negative ( Fig.~\ref{fig:fig1} (a,b)). Also, an extra peak appears in 
$\kappa_{\rm loc}$ in the topological phase (Fig.~\ref{fig:fig1}  (b)) due to the formation of an Andreev bound state as explained in main text.  
For an on-site impurity potential $V_0 > \mu^+$, close the topological phase transitions $\mu=\mu^\pm=3t$, the local particle density once again
starts to decrease as the chemical potential increases and local compressibility becomes negative (Fig.~\ref{fig:fig1} (c,d)). The additional extra peak in the compressibility now appears in the trivial phase in addition to peak in the topological phase (Fig.~\ref{fig:fig1}  (d)), once again arising because of an 
Andreev bound state.  
 
\begin{figure}
\includegraphics[width=8cm]{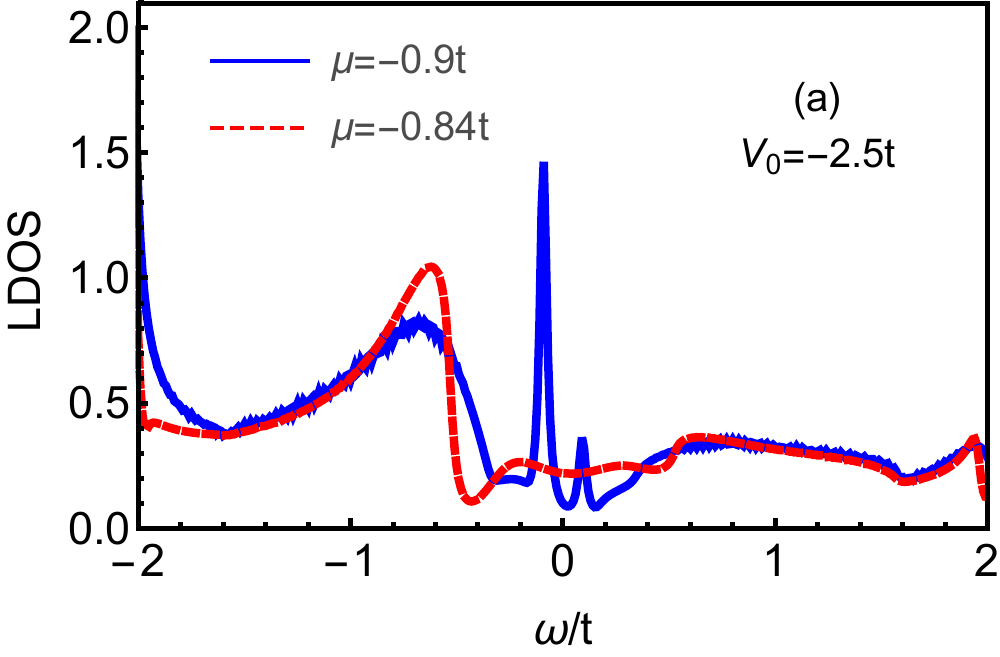}
\caption{(Color online). Formation of the bound state close to the topological phase transition for a negative on-site potential. }
\label{fig:fig2}
\end{figure}

As discussed in the main text, we discussed the role of an Andreev the bound in the presence of an on-site impurity potential 
and how that results in a decrease of the particle density with increasing $\mu$.  This was shown previously in the LDOS for a positive potential $V_0>0$. 
Here, we present the results for the LDOS in the presence of a negative potential $V_0<0$. In the latter case, the local particle density starts to {\it increase} as the $\mu$ increases because the bound state now forms {\it below} the Fermi level so more states shift below the Fermi level (Fig.~\ref{fig:fig2}).  Thus in contrast to the scenario of the positive impurity potential, the bound state {\it does} contribute to the local particle density for a negative on-site impurity.

\bibliographystyle{apsrev4-1}
\bibliography{b}